\documentclass[preprint,12pt]{elsarticle}

\usepackage{graphicx}
\usepackage{pdfpages}
\usepackage{subfigure}
\usepackage{float}
\usepackage{lineno}
\usepackage{amsmath}
\usepackage{multirow}

\journal{Nuclear Instruments and Methods in Physics Research A}

\biboptions{sort&compress}

\begin{document}

\begin{frontmatter}



\title{Developing a $\mu$Bq/m$^{3}$ level $^{226}$Ra concentration in water measurement system for the Jiangmen Underground Neutrino Observatory}

\renewcommand{\thefootnote}{\fnsymbol{footnote}}
\author{
C.~Li$^{a}$,
B.~Wang$^{a}$,
Y.~Liu$^{b}$,
C.~Guo$^{c,d,e}\footnote{Corresponding author. Tel:~+86-01088236256. E-mail address: guocong@ihep.ac.cn (C.~Guo).}$,
Y.P.~Zhang$^{c,d,e}$,
J.C.~Liu$^{c,d,e}$,
Q.~Tang$^{a}\footnote{Corresponding author. Tel:~+86-13974753537. E-mail address: tangquan528@sina.com (Q.~Tang).}$,
T.Y.~Guan$^{a}$,
C.G.~Yang$^{c,d,e}$
}
\address{
${^a}$ School of Nuclear Science and Technology, University of South China, Hengyang, 421001, China\\
${^b}$ Advanced Research Institute, Chinese Academy of Sciences, Shanghai, 201210, China\\
${^c}$ Experimental Physics Division, Institute of High Energy Physics, Chinese Academy of Sciences, Beijing, 100049, China \\
$^{d}$ School of Physics, University of Chinese Academy of Sciences, Beijing, 100049, China \\
$^{e}$ State Key Laboratory of Particle Detection and Electronics, Beijing, 100049, China
}



\begin{abstract}
The Jiangmen Underground Neutrino Observatory (JUNO), a 20~kton multi-purpose low background Liquid Scintillator (LS) detector, was proposed primarily to determine the neutrino mass ordering. To suppress the radioactivity from the surrounding rocks and tag cosmic muons, the JUNO central detector is submerged in a Water Cherenkov Detector (WCD). In addition to being used in the WCD, ultrapure water is used in LS filling, for which the $^{226}$Ra concentration in water needs to be less than 50~$\mu$Bq/m$^3$. To precisely measure the $^{226}$Ra concentration in water, a 6.0~$\mu$Bq/m$^3$ $^{226}$Ra concentration in water measurement system has been developed. In this paper, the detail of the measurement system as well as the $^{226}$Ra concentration measurement result in regular EWII ultrapure water will be presented. 

\end{abstract}


\begin{keyword}



JUNO\sep Ultrapure water\sep $^{226}$Ra\sep Mn-fiber\sep $^{222}$Rn

\end{keyword}

\end{frontmatter}


\section{Introduction}
The Jiangmen Underground Neutrino Observatory (JUNO) is a state-of-the-art liquid scintillator (LS) based neutrino physics experiment under construction in South China. Thanks to the 20~ktons of ultra-pure LS, JUNO can perform innovative and groundbreaking measurements like determining neutrino mass ordering (NMO). Beyond NMO, JUNO will measure the three neutrino oscillation parameters with a sub-percent precision~\cite{r1}. Moreover, the JUNO experiment is also expected to have important physics reach with solar neutrinos, supernova neutrinos, geoneutrinos, and atmospheric neutrinos, and searches for physics beyond the Standard Model such as nucleon decay~\cite{r2}. The JUNO detector is being constructed in a 700~m underground laboratory (1800 m.w.e.), about 53 km from the Taishan and Yangjiang nuclear power plants. JUNO will use 17,612 20-inch and 25,600 3-inch photomultiplier tubes (PMTs) to detect the light emitted from the LS. To suppress the radioactivity from the surrounding rocks and tag cosmic muons, the JUNO central detector (CD) is submerged in a Water Cherenkov Detector (WCD), which consists of 35~kton of ultrapure water and 2400 20-inch MicroChannel Plate PMTs (MCP-PMTs). In consideration of background and safety requirements, the CD and WCD must be filled with ultrapure water simultaneously. Subsequently, the ultrapure water in the CD is replaced with LS in a process known as LS filling. During replacement, the LS will be in direct contact with the ultrapure water. According to the background estimation~\cite{r1}, to avoid the ultrapure water from contaminating the LS, the $^{226}$Ra concentration in the ultrapure water has to be reduced to less than 50~$\mu$Bq/m$^3$.

$^{226}$Ra has a half-life of $\sim$1600 years, and 50~$\mu$Bq/m$^3$ is equivalent to $\sim$2.5 $\times$ 10$^{-21}$~g/g of specific activity. Such a low concentration of $^{226}$Ra can not be detected using conventional detection methods such as high-purity germanium detectors or ICP-MS. The measurement method used in this work, which was pioneered and developed by the SNO collaboration~\cite{r3,r4}, is to first enrich and extract the $^{226}$Ra from water using MnO$_x$, and then determine the concentration of $^{226}$Ra in water by its gaseous daughter $^{222}$Rn. Different from the SNO experiment, we used a new low-background large surface area material, polyurethane fibers, as the carrier of MnO$_x$, a highly sensitive radon emanation measurement device to measure the $^{226}$Ra activity, and a standard $^{226}$Ra solution to calibrate the radium extraction efficiency. In addition, we also used the relative measurement to exclude the effect of pre-existing $^{226}$Ra ions in the ultrapure water on the extraction efficiency.

This paper is organized as follows. Sec.2 describes the $^{226}$Ra measurement principle, sec.3 describes the experimental setup, sec.4 describes the calibration of the extraction efficiency, the estimation of measurement sensitivity as well as the measurement result of $^{226}$Ra concentration in ultrapure water, and sec.5 is the summary.

\section{Measurement principle}

\begin{figure}[htb]
	\centering
    \includegraphics[width=6cm]{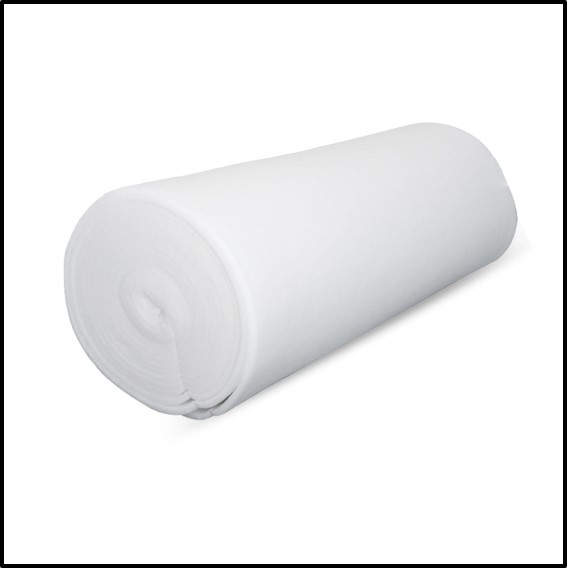}
	\includegraphics[width=4.5cm]{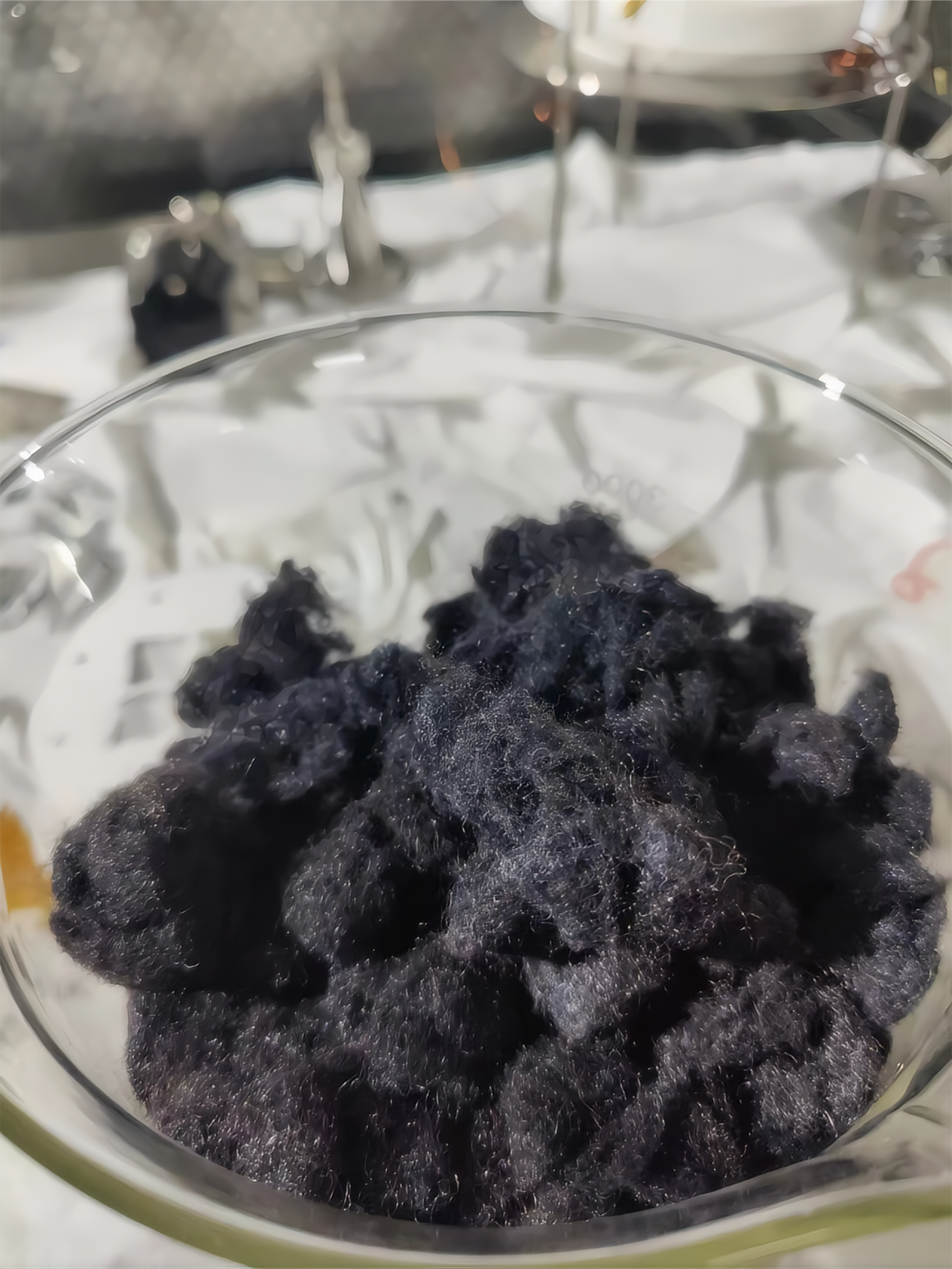}
	\caption{Pictures of polyurethane fiber (left) and Mn-fiber (Right).}
	\label{fig.fibers}
\end{figure}

The initial step in measuring the $^{226}$Ra concentration in water is the extraction of $^{226}$Ra from water. According to Ref~\cite{r5,r6,r7,r8,r9}, manganese oxide compounds ~(MnO$_{x}$) have a strong adsorption capability for $^{226}$Ra. The material used in this work to extract $^{226}$Ra from water is Mn-fiber, which is polyurethane fiber attached with MnO$_{x}$ powder. The images of the polyurethane fiber and Mn-fiber are depicted in Fig.~\ref{fig.fibers}. The manufacturing process of Mn-fiber will be elaborated upon in Section 3.

\begin{figure}[htb]
	\centering
	\includegraphics[width=6.5cm]{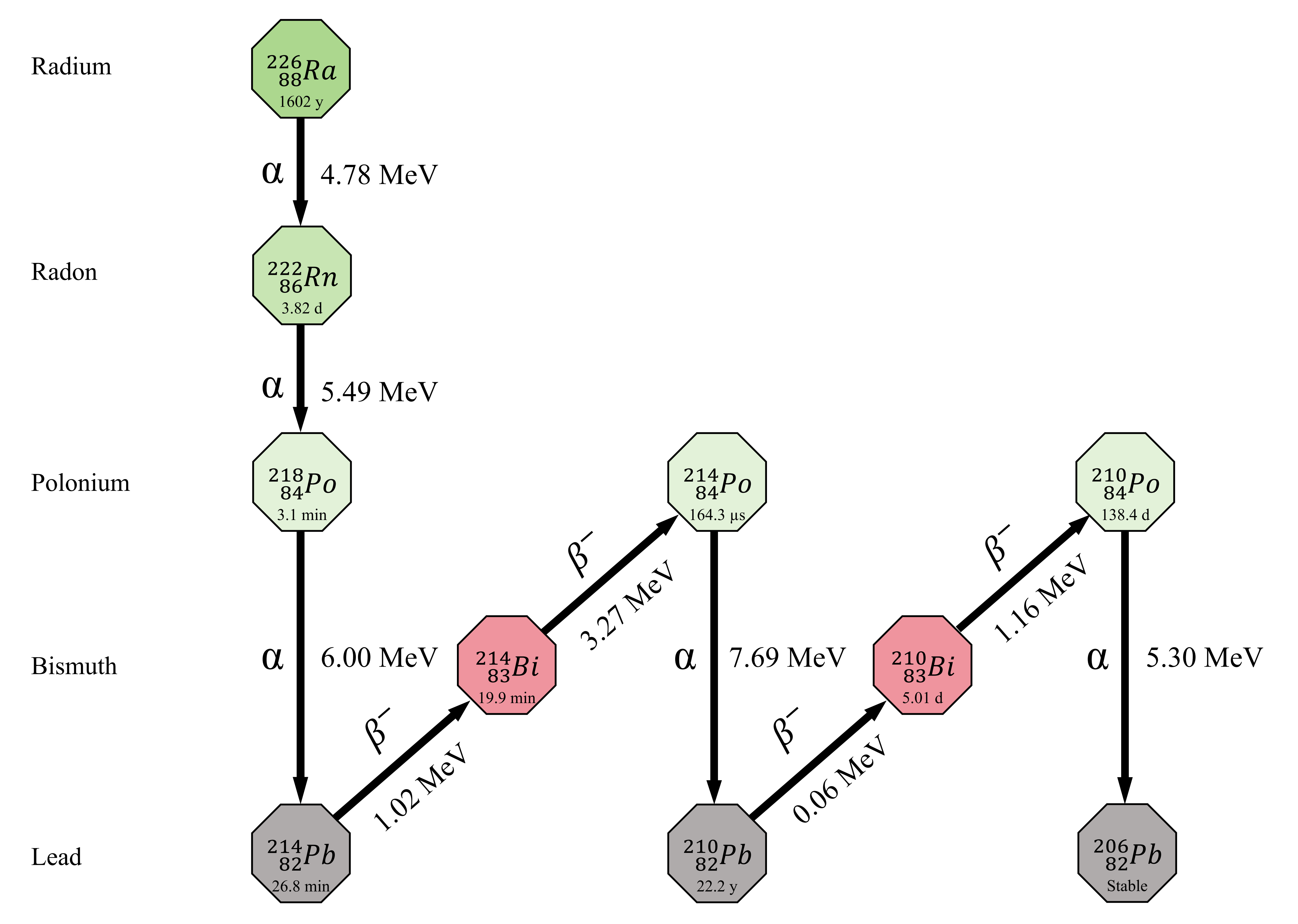}
	\caption{The relevant decay branches of $^{226}$Ra. The times below the elements are the half-lives. The energy for $\alpha$ decay is the peak energy while for $\beta$ decay is the maximum energy.}
	\label{fig.decaychain}
\end{figure}

After extracting $^{226}$Ra from water, the subsequent step is determining the $^{226}$Ra activity. A radon emanation measurement system is utilized for this purpose. The relevant decay branches of $^{226}$Ra are illustrated in Fig.~\ref{fig.decaychain}. As the precursor of $^{222}$Rn, the activity of $^{226}$Ra can be ascertained by measuring the activity of $^{222}$Rn. Before the measurement, the Mn-fiber, which has been utilized for the extraction of $^{226}$Ra from water, has to be hermetically sealed in a container to allow $^{226}$Ra to decay and produce $^{222}$Rn~\cite{r10}. Given the significantly longer half-life of $^{226}$Ra compared to that of $^{222}$Rn, the relationship between the activities of $^{226}$Ra and $^{222}$Rn can be expressed by Eq.~\ref{Eq.1}:
\begin{equation} 
A_{Ra} = \frac{A_{Rn}}{1-e^{-\lambda _{Rn}t}} 
\label{Eq.1}
\end{equation}
where A$_{Ra}$ is the activity of $^{226}$Ra in the unit of mBq, A$_{Rn}$ is the activity of $^{222}$Rn in the unit of mBq, $\lambda_{Rn}$ is the decay constant of $^{222}$Rn, and t is the sealing time of Mn-fiber in the unit of day. The details of the radon emanation measurement will be described in Sec.3.2. 

\section{Experimental setup}
The $^{226}$Ra concentration in water measurement system consists of a $^{226}$Ra extraction column and a set of radon emanation measurement system. The extraction column is used to extract $^{226}$Ra when water flows through and the emanation system is used to determine the $^{226}$Ra activity.

\subsection{Extraction column}
\begin{figure}[htb]
	\centering
    \includegraphics[height=5cm]{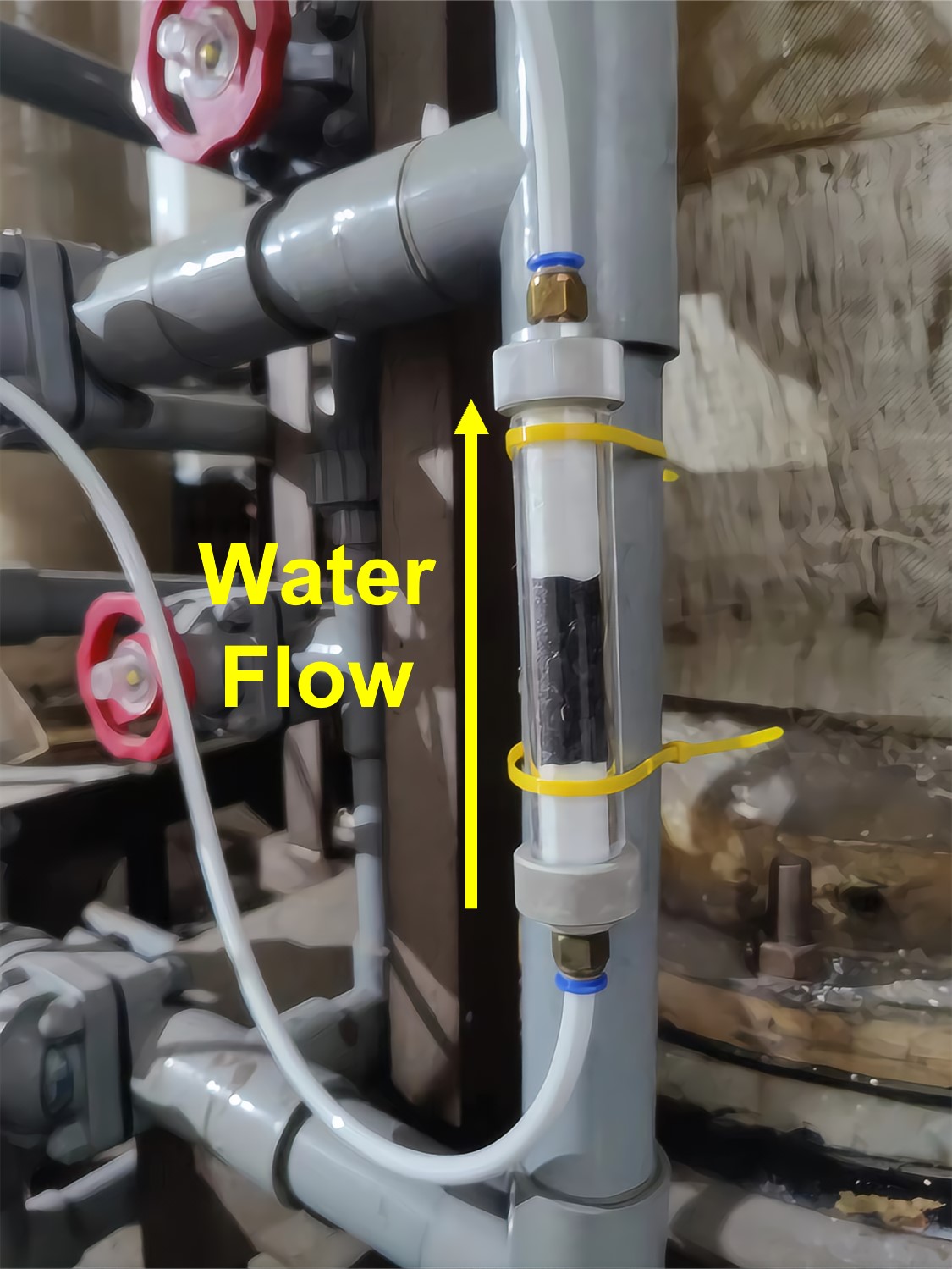}
	\caption{A picture of the extraction column. The black part in the middle is Mn-fibers and the white parts at both ends are polyurethane fibers. Each column contains $\sim$5~g of Mn-fiber. The primary function of the polyurethane fibers is to ensure dense packing within the column, facilitating full and uniform contact with the Mn-fibers by the water.}
	\label{fig.column}
\end{figure} 
As is shown in Fig.~\ref{fig.column}, the extraction column is an acrylic tube with rubber gaskets at both ends, which can withstand about 5~kg of water pressure. The extraction column has an inner diameter of about 20~mm and a length of about 250~mm. The main components of the extraction column are the Mn-fibers located in the middle, and the upper and lower parts of the column are filled with white fibers, whose main function is to filter out any tiny particles that may be present in the water. To make full contact between the water and the Mn-fibers, the water flows upwards from the bottom. For each measurement, $\sim$5~g of Mn-fibers are used to fill about half of the column.

The Mn-fibers are made from polyurethane fibers, potassium permanganate (KMnO$_4$) solution, and concentrated sulfuric acid. The manufacturing processes can be found in our previous work~\cite{r11}, while in this work we have made the following improvements:

(A) The raw materials are carefully screened. The $^{226}$Ra backgrounds of polyurethane fibers and KMnO$_4$ powder are measured using the radon emanation system and the backgrounds for those used in this work are shown in Tab.~\ref{tab.rawmaterial}. In addition, the purity of the concentrated sulfuric acid used in the experiment was upgraded from analytically pure to electronic grade. 

\begin{table}[htb]
\centering
\setlength{\tabcolsep}{4.5mm}{
\caption {The $^{226}$Ra concentrations of polyurethane fibers and KMnO$_4$ powder used in this work.}
\begin{tabular}{ccc}
\hline
Material & Sample weight (g) & $^{226}$Ra concentration ($\mu$Bq/g)   \\
\hline
Polyurethane fiber &   75  &  1.6 $\pm$ 1.4 \\
KMnO$_4$ powder &  250  &  0.65 $\pm$ 0.52  \\
\hline
\label{tab.rawmaterial}	
\end{tabular}}
\end{table}

(B) The container used to make Mn-fibers was changed from a glass beaker to a quartz beaker to prevent contamination of the Mn-fibers by impurities in the glass beaker in a strongly acidic and hot environment.

(C) The polyurethane fibers were cut into small pieces of $\sim$1~cm $\times$ 1~cm, which allowed them to be denser when filled into the extraction column, resulting in a more even and full contact with the water while extracting $^{226}$Ra.

(D) While rinsing the Mn-fibers, a shaking machine is used to help remove the inefficiently adsorbed MnO$_x$ on the fiber, which reduces the shedding amount of MnO$_x$ during extraction.

According to the background measurement and the $^{226}$Ra extraction efficiency calibration results, the background of Mn-fibers was reduced to $\sim$17~$\mu$Bq/g and the extraction efficiency was increased to $\sim$95\% after these improvements. Details about the extraction efficiency calibration will be discussed in Sec.4.

\subsection{Radon emanation measurement system}

\begin{figure}[htb]
	\centering
	\includegraphics[height=5cm]{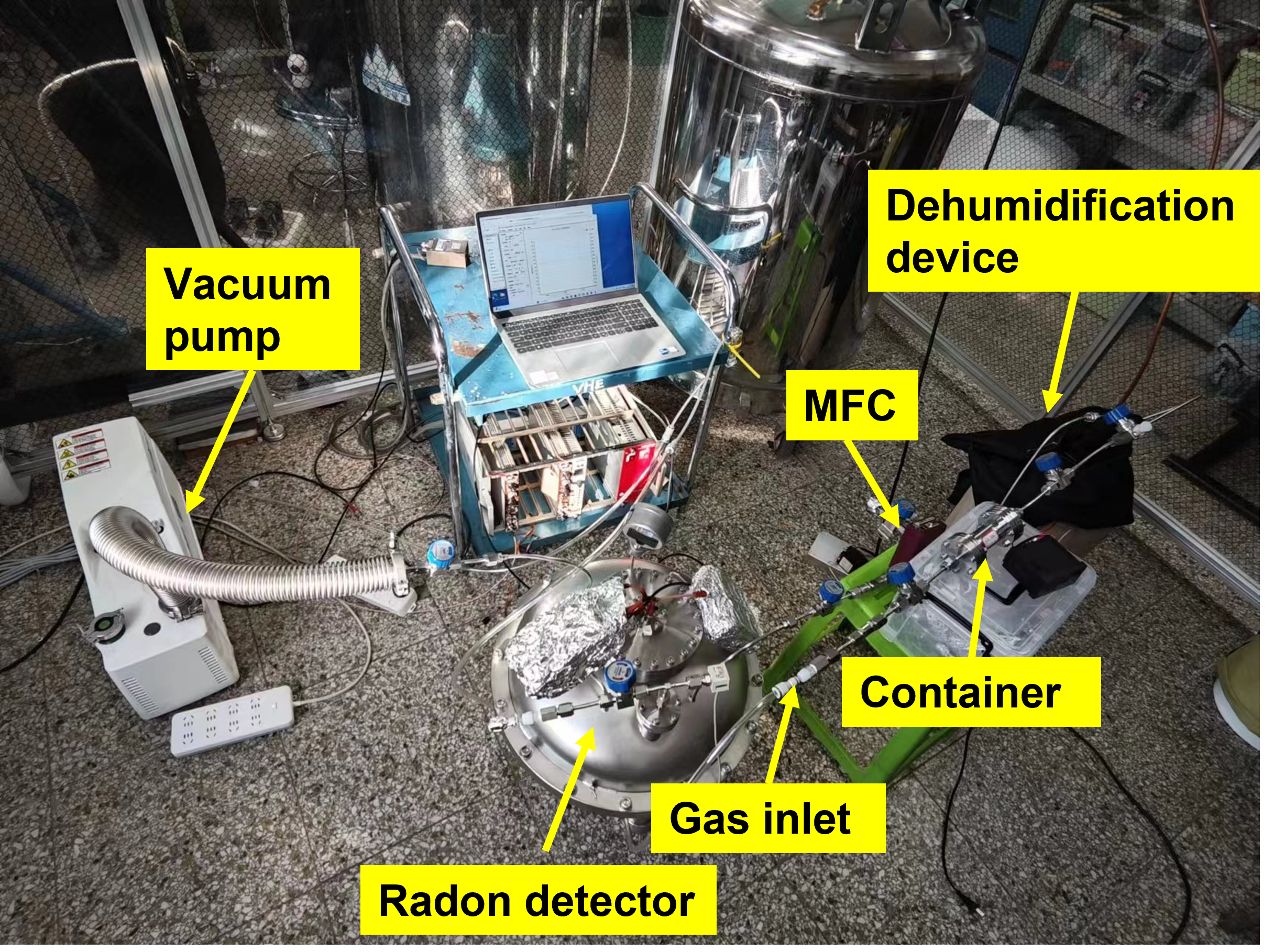}
    \includegraphics[height=5cm]{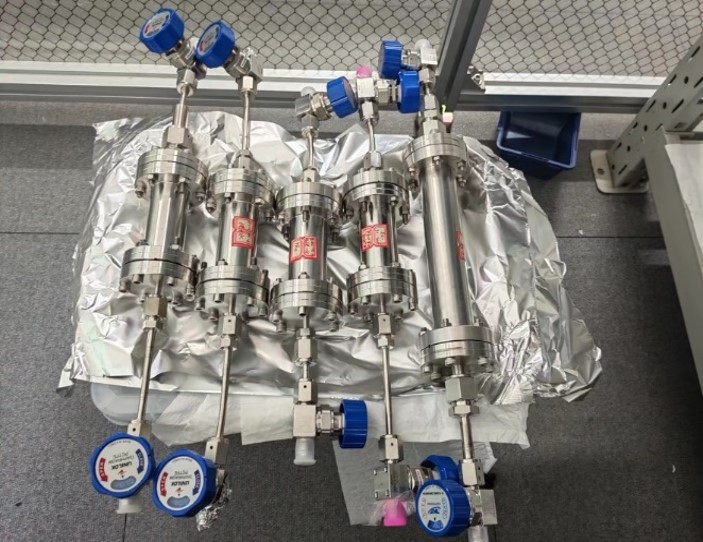}
	\caption{Left: A picture of the radon emanation measurement system. The system consists of a container, a dehumidification device, an MFC, a radon detector, and a vacuum pump. The container is used for Mn-fiber sealing, the dehumidification device is used to reduce the humidity of the gas, the MFC is used to control the gas flow rate, the radon detector is used to determine the radon activity in the gas, and the vacuum pump is used to vacuum the detector before gas transfer. The gas inlet is connected to the vent of a liquid nitrogen dewar, which supplies evaporating nitrogen to purge radon gas from the Mn-fiber container to the detector. Right: A picture of the containers. Multiple containers of varying sizes were used to accommodate the closure of samples for a few days, enabling the $^{226}$Ra decay to produce $^{222}$Rn for each measurement, as well as to account for difference sample masses during the experiment. }
	\label{fig.container}
\end{figure}

The $^{226}$Ra activity is determined by measuring its gaseous daughter $^{222}$Rn with a radon emanation measurement system. As is shown in the left of Fig.~\ref{fig.container}, the system consists of a Mn-fiber container, a set of dehumidification devices, a Mass Flow Controller (MFC, 1179A MKS), a vacuum pump (ACP40, Pfeiffer), and a radon detector. In this system, ConFlat (CF) flanges with metal gaskets and VCR pipelines and valves with metal gaskets are used to keep the air leakage of the system to better than 1 $\times$ 10$^{-9}$ Pa*m$^3$/s, which is measured by a helium leak detector (ZQJ-3000, KYKY Technology Co. Ltd). The details and functions of each component are as follows:

(A) Mn-fiber Container. The right of Fig.~\ref{fig.container} shows a picture of the container, which is a Stainless Steel (SS) tube with CF35 flanges at both ends. To facilitate sealing and connection to the emanation system, the connectors of the container have been converted to 1/4 VCR valves. Once the Mn-fibers are removed from the extraction column, they are packed into this container, which is purged with evaporated nitrogen at a flow rate of 1 L/min for 30 minutes before being sealed, thus eliminating the impact of air on the measurement results. The Mn-fibers are sealed in a container for several days before the measurement, allowing the $^{226}$Ra present on them to decay and produce $^{222}$Rn. 

(B) The gas inlet is connected to the vent of a high-pressure liquid nitrogen dewar, which supplies evaporating nitrogen to purge radon gas from the Mn-fiber container to the detector. 

(C) Dehumidification devices. The Mn-fiber taken out from the column retains moisture, leading to high humidity as the gas passes through the container. This elevated humidity can decrease the detection efficiency of the radon detector\cite{r12}. To address this, a dehumidifying device is employed to regulate the relative humidity of the gas to less than 3\%. The details of the dehumidification device can be found in our previous work~\cite{r11,r13}.

(D) MFC. The MFC is utilized to regulate the gas flow rate during the transfer of radon gas.

(E) Vacuum pump. The vacuum pump is utilized for gas transfer. Before each measurement, the radon detector and pipelines are evacuated with the vacuum pump, and subsequently, evaporated nitrogen is utilized to purge the gas from the container into the detector.

\begin{figure}[htb]
	\centering
    \includegraphics[height=3.1cm]{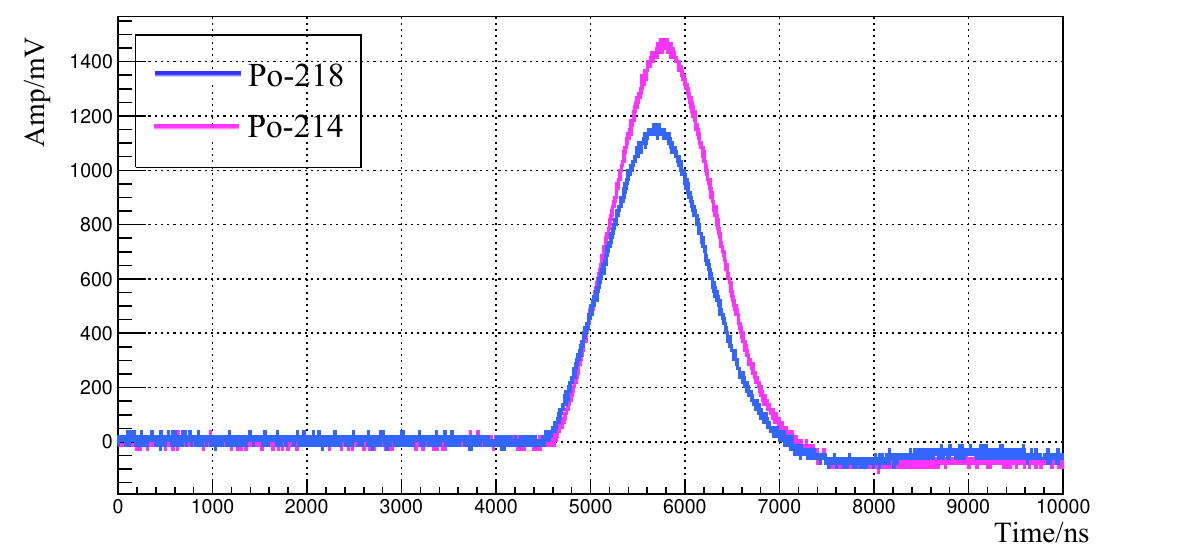}
	\includegraphics[height=3.1cm]{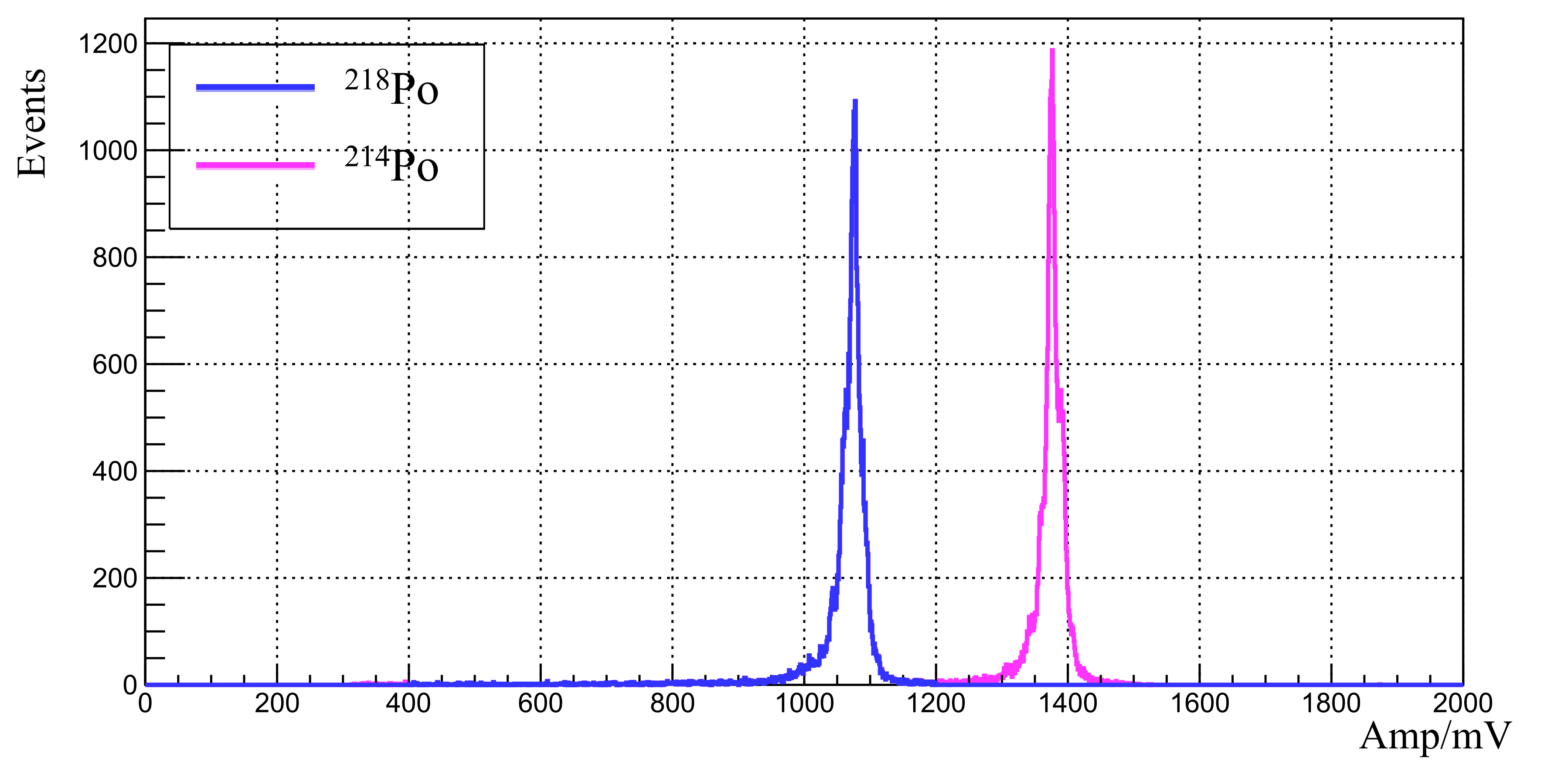}
	\caption{Left: The example pulses measured by the radon detector. Right: The amplitude spectrum measured with a $^{222}$Rn source.}
	\label{fig.examples}
\end{figure}

(F) Radon detector. The radon detector is utilized to determine the $^{222}$Rn activity. The background event rate for the detector itself is 0.70 $\pm$ 0.15 Counts Per Day (CPD) and the Calibration Factor (C$_F$) of the detector is 67.0 $\pm$ 6.7 Counts Per Hour per Bq/m$^3$ (CPH/(Bq/m$^3$)), which corresponds to a one-day measurement sensitivity of 0.71~mBq/m$^3$~\cite{r12}. The uncertainty of the background event rate is statistical while the uncertainty of C$_F$ is systematical which is caused by the variation of radon source. Example pulses and the amplitude spectrum measured by the radon detector are shown in Fig.~\ref{fig.examples}. The details of the radon detector can be found in Ref.~\cite{r12,r14,r15}.

The background event rate for the entire system, encompassing the radon detector, container, dehumidification device, pipelines, and evaporated nitrogen, is 3.0 $\pm$ 1.7 CPD. According to Eq.~\ref{Eq.2}~\cite{r16}, the sensitivity of the radon emanation measurement is 72~$\mu$Bq/sample.

\begin{equation} 
L = \frac{1.64 \times \sigma_{BG}}{C_F \times 24} 
\label{Eq.2}
\end{equation}
where L is the sensitivity, $\sigma_{BG}$ is the statistical uncertainty of the background measurement and C$_F$ is the calibration factor of the radon detector.

After the transfer of $^{222}$Rn gas into the detector, the $^{222}$Rn undergoes decay with a half-life of 3.8 days. As the measurement often extends beyond 20 hours, the detector actually records the average $^{222}$Rn concentration. The relationship between the average concentration and the initial concentration transferred to the detector can be expressed by Eq.~\ref{Eq.3}:
\begin{equation} 
A_{Rn} = \frac{A_{Rn-M} \times \lambda_{Rn} \times (t_2-t_1)}{e^{-\lambda_{Rn}t_1}-e^{-\lambda_{Rn}t_2}} 
\label{Eq.3}
\end{equation}
where A$_{Rn}$ is the initial $^{222}$Rn activity after transferring into the detector in the unit of mBq, A$_{Rn-M}$ is the measured $^{222}$Rn activity by the detector in the unit of mBq, t$_1$ is start measurement time in the unit of day, and t$_2$ is the end measurement time in the unit of day. Once the initial activity of $^{222}$Rn is determined, the activity of $^{226}$Ra can be calculated using Eq.~\ref{Eq.1}.

\section{Results}
\subsection{$^{226}$Ra solution}
A $^{226}$Ra solution, which was provided by South China University, was used to calibrate the $^{226}$Ra extraction efficiency of Mn-fiber from the water. The $^{226}$Ra concentration in the solution was measured by the radon emanation measurement system. 3~mL of $^{226}$Ra solution was evenly distributed onto a polyurethane fiber, which is $\sim$10~cm in length and $\sim$2~cm in width, and subsequently placed in a Mn-fiber container. The Mn-fiber container was purged with evaporated nitrogen gas at a flow rate of 1~L/min for 30 minutes before the valves at both ends were closed. After several days of sealing, the radon activity was measured using the radon emanation measurement system. Then the $^{226}$Ra activity in the radium solution was calculated according to Eq.~\ref{Eq.3} and Eq.~\ref{Eq.1}. According to the measurement results, the $^{226}$Ra concentration in the solution is 9.1 $\pm$ 1.1~mBq/mL, the uncertainty consists of both systematic and statistical components.

\subsection{The ultrapure water system}

\begin{figure}[htb]
	\centering
    \includegraphics[height=5cm]{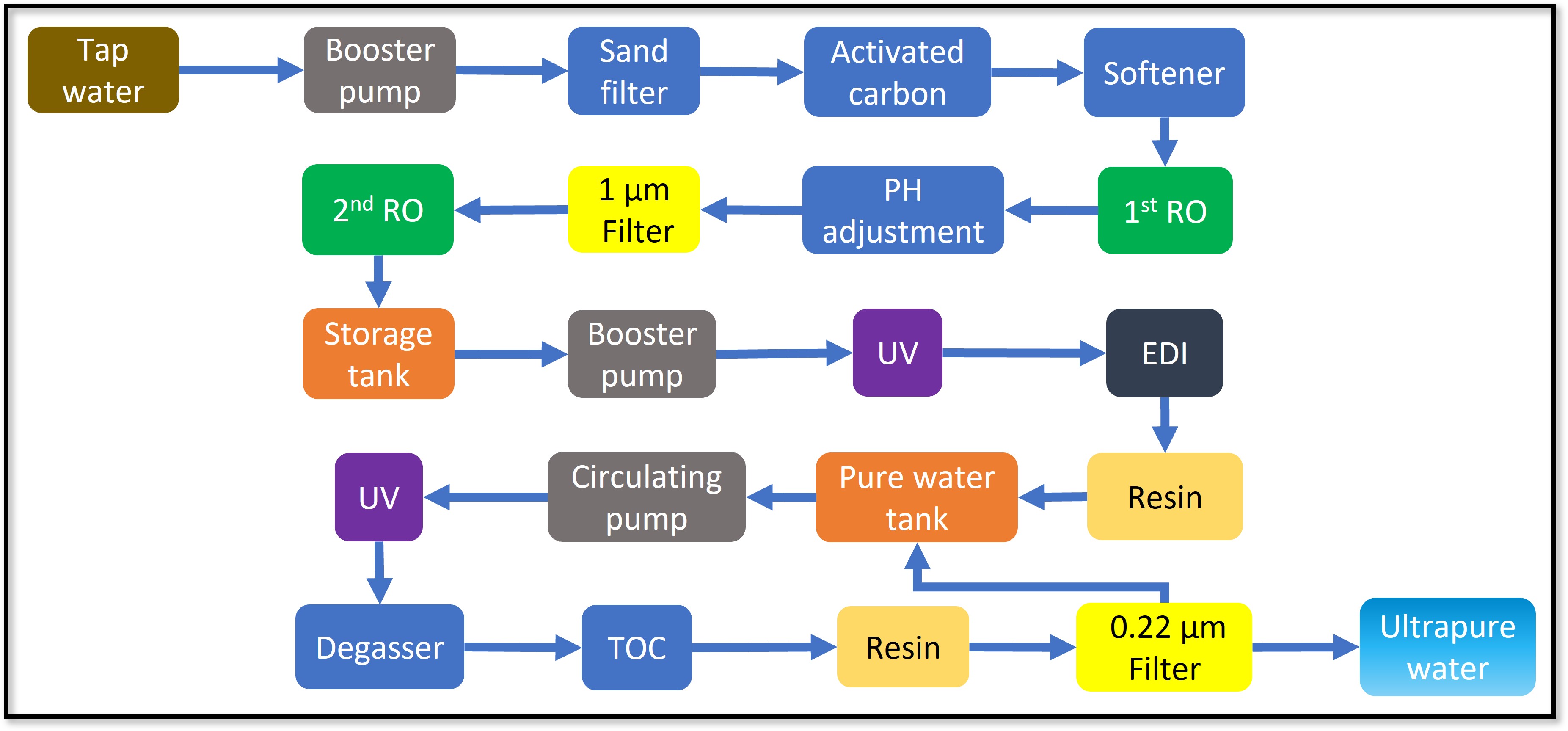}
	\caption{The flow chart of the water system. }
	\label{fig.watersystem}
\end{figure} 

The ultrapure water utilized in this experiment was supplied by the ultrapure water station of the Institute of High Energy Physics (IHEP). The water system is designed for EWII-class ultrapure water, and Fig.~\ref{fig.watersystem} illustrates the flow chart of the water system. Radium primarily exists in the form of ions in water. Based on the working principles of each device in the water system, the sand filter, RO, EDI, and resin can remove radium. The system can work in two modes automatically according to the water consumption, one is water production mode and the other is circulation mode. In the recirculation mode, the produced ultrapure water is returned to the pure water tank, allowing the water to repeatedly pass through the resin. As a result, the radium concentration gradually decreases over time.

\subsection{Extraction efficiency calibration}
\begin{figure}[htb]
	\centering
    \includegraphics[height=5cm]{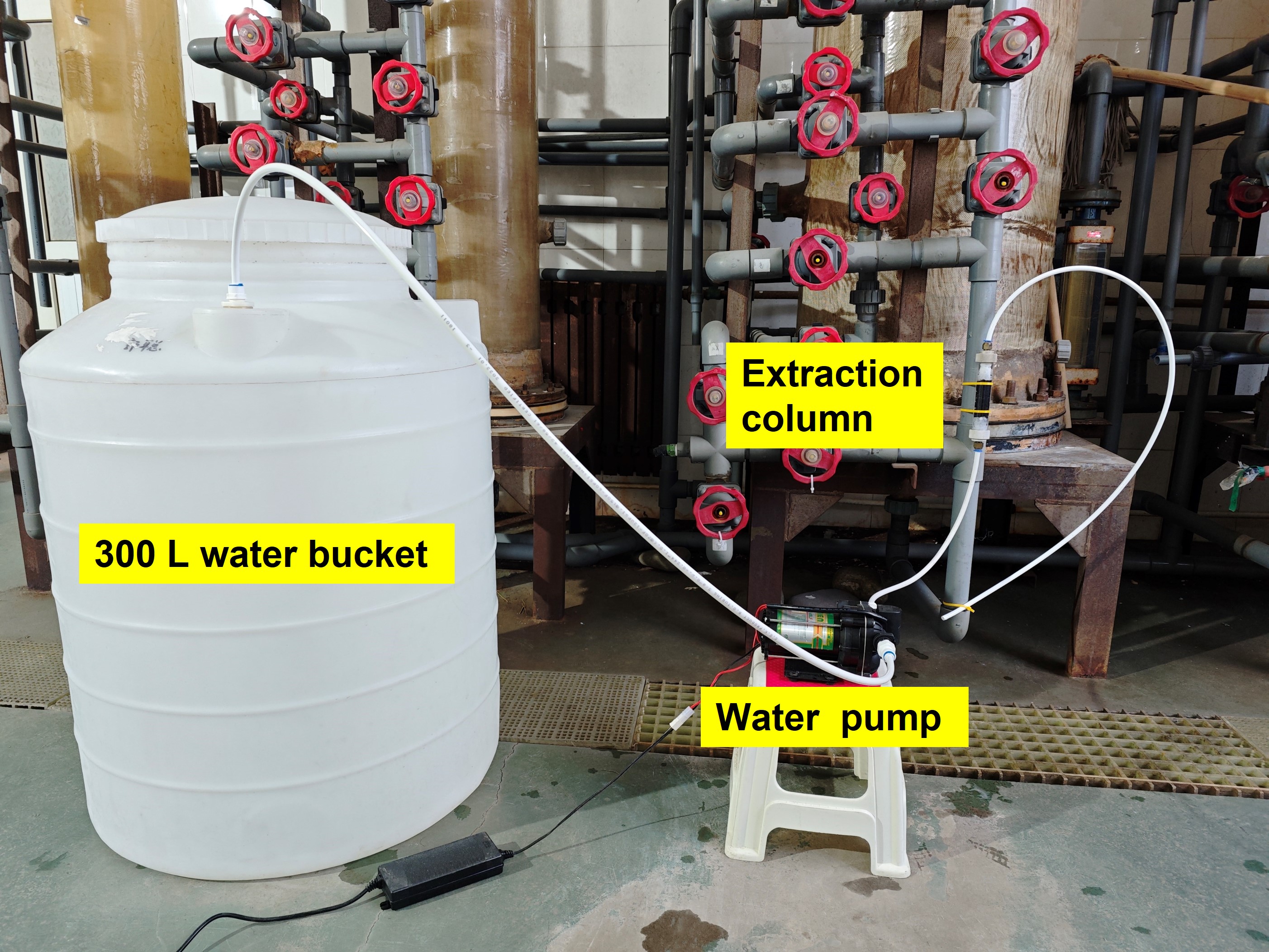}
	\caption{Picture of the $^{226}$Ra extraction efficiency measurement system. }
	\label{fig.rameasurement}
\end{figure}

When conducting measurements of radium extraction efficiency, we utilized relative measurements. Two identical systems were employed, consisting of the same volume of pure water buckets, the same type of pumps, identical size adsorption columns, and the same mass of Mn-fibers and polyurethane fibers. Fig.~\ref{fig.rameasurement} shows a picture of one set of the measurement system. Radium solution was added for one set of experiments, while it was not added for the other. The advantage of using relative measurements is that the original radium present in the ultrapure water can be circumvented from influencing the measurement results. In addition, relative measurements counteract the systematic uncertainty associated with the detector itself.

While assessing the factors that may affect the $^{226}$Ra extraction efficiency of Mn-fibers, our primary considerations are the $^{226}$Ra concentration in water and the water flow rate.

\subsubsection{Effect of $^{226}$Ra concentration in water on the extraction efficiency}
To investigate the effect of $^{226}$Ra concentration in water on the adsorption efficiency, the experiments were conducted using aqueous $^{226}$Ra solutions with concentrations ranging from 0.76~mBq/m$^3$ to 61~mBq/m$^3$. During the tests, the volume of water remained constant at 300~L, with a flow rate of 5~L/min through the manganese fibers. Approximately 5~g of Mn-fibers and 5~g of polyurethane fibers were utilized in each experiment. For the control experiment, the remaining experimental setup was identical to it except that no radium solution was added to the bucket. After substructing the contribution from water, the $^{226}$Ra extraction efficiencies at various $^{226}$Ra concentrations in water are shown in Tab.~\ref{tab.differentC}. The uncertainties of the added $^{226}$Ra activity and the extraction efficiency consist of both systematic and statistical uncertainties, while the uncertainties of the extracted $^{226}$Ra activity are statistical only. No variation in adsorption efficiency with concentration was observed within the range of concentrations used in this experiment. 

\begin{table}[htb]
	\centering
	\setlength{\tabcolsep}{1.8mm}{
		\footnotesize
		\caption {$^{226}$Ra extraction efficiencies with various $^{226}$Ra concentrations in water }
		\begin{tabular}{cccc}
			\hline
			Sample & Added $^{226}$Ra activity (mBq) & Extracted $^{226}$Ra activity (mBq) & Extraction efficiency (\%)   \\\hline 
			1 &     0.228 $\pm$ 0.028  & 0.21 $\pm$ 0.19  &  92 $\pm$ 84  \\
			2 &     0.455 $\pm$ 0.055  & 0.43 $\pm$ 0.24  &  95 $\pm$ 54  \\		
			3 &     0.91  $\pm$ 0.11   & 0.83 $\pm$ 0.25  &  91 $\pm$ 30  \\
			4 &     1.82  $\pm$ 0.22   & 1.68 $\pm$ 0.32  &  92 $\pm$ 21  \\			
			5 &     4.55  $\pm$ 0.55   & 4.40  $\pm$ 0.51  &  97 $\pm$ 16  \\
			6 &     18.2  $\pm$ 2.2    & 17.1 $\pm$ 1.0   &  94 $\pm$ 13  \\\hline 
			\label{tab.differentC}	
	\end{tabular}}
\end{table}

\subsubsection{Effect of water flow rate on the extraction efficiency}
When testing the impact of water flow rate on the $^{226}$Ra extraction efficiency, three different flow rates were used: 1.6~L/min, 5~L/min, and 8~L/min. Apart from this, the conditions were consistent across all three tests: a volume of 300 L of ultrapure water, 1.5~mL of $^{226}$Ra solution, 5~g of Mn-fibers, 5~g of polyurethane fibers, as well as control experiments without added $^{226}$Ra solution. The results are shown in Tab.~\ref{tab.differentF}. The uncertainties for added $^{226}$Ra activity and extraction efficiency consist of both systematic and statistical uncertainties, while the uncertainties of the extracted  $^{226}$Ra activity are statistical only. These results also indicate that increasing the water flow rate to 8~L/min has minimal effect on the extraction efficiency of $^{226}$Ra from water by Mn-fibers.

\begin{table}[htb]
	\centering
	\setlength{\tabcolsep}{1mm}{
		\footnotesize
		\caption {$^{226}$Ra extraction efficiencies with various water flow rate}
		\begin{tabular}{cccc}
			\hline
			Flow rate (L/min) & Added $^{226}$Ra activity (mBq) & Extracted $^{226}$Ra activity (mBq) & Extraction efficiency (\%)   \\\hline 
			1.6 & 13.7 $\pm$ 1.7 & 12.8 $\pm$ 0.73 &  93 $\pm$ 10     \\
			5.0 & 13.7 $\pm$ 1.7 & 12.7 $\pm$ 0.90 &  93 $\pm$ 13          \\		
			8.0 & 13.7 $\pm$ 1.7 & 12.5 $\pm$ 0.78 &  91 $\pm$ 13         \\\hline 
			\label{tab.differentF}	
	\end{tabular}}
\end{table}

\subsection {Sensitivity estimation}
The sensitivity of $^{226}$Ra concentration in water measurement at 90\% confidence level can be estimated according to Eq~\ref{eq.sensitivity}:

\begin{equation}
L\textquotesingle=\frac{1.64 \times \sigma_{bg}\textquotesingle \times V_s}{24 \times C_F \times V_w \times \varepsilon \times 10^{-6}} 
\label{eq.sensitivity}
\end{equation}
where L$\textquotesingle$ is the sensitivity in the unit of $\mu$Bq/m$^3$, $\sigma_{bg}\textquotesingle$ is the statistical uncertainty of the total background, V$_D$ is the volume of the whole system which is 0.042~m$^3$, C$_F$ is the calibration factor of the radon detector, V$_w$ is the volume of water treated by the Mn-fibers in the unit of m$^3$ and $\varepsilon$ is the $^{226}$Ra extraction efficiency. The total background of the system can be calculated according to Eq.~\ref{eq.background}:

\begin{equation}
	N_{bg}=\frac{m \times A_{bg}}{V_s}\times C_F \times 24 + n_s 
	\label{eq.background}
\end{equation}
where N$_{bg}$ is the background event rate of the whole system, including the Mn-fiber and the radon emanation measurement system, in the unit of CPD, m is the mass of Mn-fibers used in one measurement in the unit of g, A$_{bg}$ is the Mn-fiber $^{226}$Ra concentration in the unit of mBq/g, n$_s$ is the background event rate of the radon emanation measurement system in the unit of CPD.

Before determining the final sensitivity, it is necessary to establish the water treatment capacity of a unit mass of Mn-fiber. According to the SNO experiment measurements, approximately 1g of MnO$_x$ can be utilized to extract $^{226}$Ra from approximately 46~m$^3$ of ultrapure water with an efficiency of $\sim$95\%~\cite{r4,r17}. Based on the mass measurements of Mn-fibers and polyurethane fibers, we found that the mass of MnO$_x$ attached to the Mn-fibers accounts for approximately 9\% of the total mass. This suggests that 5~g of Mn-fibers contain roughly 0.45~g of MnO$_x$. Therefore, theoretically, 5g of Mn-fibers can extract $^{226}$Ra from approximately 20~m$^3$ of water. In this experiment, we employed a radium solution to assess the extraction efficiency of radium adsorption in 20~m$^3$  of water by 5~g of Mn-fibers. Based on the measurement results shown in Tab.~\ref{tab.15m3}, it is determined that 5g of manganese fibers can adsorb radium in 20~m$^3$ of water with an extraction efficiency of $\sim$89\%.

\begin{table}[htb]
	\centering
	\setlength{\tabcolsep}{1.8mm}{
		\footnotesize
		\caption {$^{226}$Ra extraction efficiency results from 20~m$^{3}$ water by 5~g of Mn-fibers.}
\begin{tabular}{ccc}
	\hline
	Added $^{226}$Ra activity (mBq)& Extracted $^{226}$Ra activity (mBq)  & Enrichment efficiency (\%)  \\\hline 		
    82.1 $\pm$ 9.8 & 92 $\pm$ 11 & 89 $\pm$ 15 \\\hline
	\label{tab.15m3}	
\end{tabular}}
\end{table}

Based on the measurement and calibration results provided earlier, we can ascertain that the sensitivity of this system for measuring radium concentration in water is $\sim$6.0~$\mu$Bq/m$^3$. The relevant parameter values used for estimating the sensitivity are detailed in Tab.~\ref{tab.sensitivity}.

\begin{table}[htb]
	\centering
	\setlength{\tabcolsep}{1.8mm}{
		\footnotesize
		\caption {Parameters used for sensitivity estimation.}
\begin{tabular}{ccccccc}
	\hline
	m (g) & A$_{bg}$ ($\mu$Bq/g) & V$_s$ (m$^3$) & C$_F$ (CPH/(Bq/m$^3$)) &  n$_s$ (CPD) & V$_w$ (m$^3$) & $\varepsilon$ (\%)  \\\hline 		
	 5 & 17.1 $\pm$ 4.0  & 0.042  & 67.0 $\pm$ 6.7 & 3 & 15 &  89 $\pm 15$ \\\hline  
	\label{tab.sensitivity}	
\end{tabular}}
\end{table}

\subsection{$^{226}$Ra concentration in water}
In the aforementioned tests, each experimental set included a control group comprising direct measurement results of $^{226}$Ra concentration in ultrapure water. Additionally, periodic sampling and measurement of the ultrapure water were conducted throughout the experiment. As is shown in Tab.\ref{tab.water}, the $^{226}$Ra concentration in the water fluctuates within the range of approximately $\sim$140~$\mu$Bq/m$^3$ to $\sim$830~$\mu$Bq/m$^3$. The variation is attributed to the operational mode of the water system and the number of cycles of ultrapure water in the circulation mode. During the water production mode, tap water sequentially passes through each component, and due to the relatively constant $^{226}$Ra removal efficiency of each component, the $^{226}$Ra concentration in water is relatively higher in this mode. Conversely, in the circulation mode, the ultrapure water passes through the resin multiple times, and the resin effectively removes radium from the water. Consequently, the more cycles the water undergoes through the resin, the lower the $^{226}$Ra concentration in the water. Furthermore, it is important to note that the resin is consumable, and its radium removal efficiency gradually decreases with continuous use. According to the data shown in Tab.~\ref{tab.water}, for a standard EWII water system, the $^{226}$Ra concentration in the water typically ranges around one hundred to several hundred $\mu$Bq/m$^3$.

\begin{table}[htb]
	\centering
	\setlength{\tabcolsep}{1.8mm}{
		\footnotesize
		\caption {The measurement results of $^{226}$Ra concentration in ultrapure water.}
\begin{tabular}{cccc}
	\hline
	Sample & Extracted $^{226}$Ra activity ($\mu$Bq)  & Water volume (m$^3$) & $^{226}$Ra concentration ($\mu$Bq/m$^3$)  \\\hline 		
     1  & 173 $\pm$ 125 & 0.3 & 577 $\pm$ 417\\ 
     2  & 182 $\pm$ 128 & 0.3 & 607 $\pm$ 427 \\
     3  & 241 $\pm$ 168 & 0.3 & 803 $\pm$ 566\\ 
     4  & 219 $\pm$ 182 & 0.3 & 730 $\pm$ 607\\ 
     5  & 250 $\pm$ 170 & 0.3 & 833 $\pm$ 566\\ 
     6  & 308 $\pm$ 187 & 1.5 & 205 $\pm$ 125\\ 
     7  & 213 $\pm$ 167 & 1.5 & 140 $\pm$ 112 \\
     8  & 326 $\pm$ 196 & 1.5 & 217 $\pm$ 131 \\
     9  & 449 $\pm$ 217 & 1.5 & 299 $\pm$ 145 \\
	 10 & 8460 $\pm$ 1074 & 20 & 423 $\pm$ 54 	\\\hline 
	\label{tab.water}	
\end{tabular}}
\end{table}

\section{Summary}
 JUNO, a cutting-edge LS-based neutrino physics experiment currently under construction in South China, aims to determine the NMO with 3-4~$\sigma$ in 6 years. To mitigate the radioactive background from the surrounding rocks as well as tag cosmic muons, the LS detector will be submerged in a WCD, of which the primary component is 35~kton of ultrapure water. In addition, $\sim$25~kton ultrapure water will be utilized for LS filling, which requires the $^{226}$Ra concentration to be less than 50~$\mu$Bq/m$^3$. To precisely measure the $^{226}$Ra concentration in water, a system based on $^{226}$Ra extraction using Mn-fiber and  $^{226}$Ra activity determination through radon emanation has been established. The system's sensitivity, as determined from background measurement and efficiency calibration, is $\sim$6.0~$\mu$Bq/m$^3$. Using this system, we measured the $^{226}$Ra concentration in the produced water of a standard EWII water system, observing variations ranging from one hundred to several hundred $\mu$Bq/m$^3$, which were dependent on the operation mode, operation time and material consumption.

\section{Acknowledgement}
This work is supported by the State Key Laboratory of Particle Detection and Electronics (Grant No. SKLPDE-ZZ-202304), the Youth Innovation Promotion Association of the Chinese Academy of Sciences (Grant No. 2023015), and the Strategic Priority Research Program of the Chinese Academy of Sciences (GrantNo.XDA10011200).

\end{document}